# Cometary Dust Trail Associated with Rosetta Mission Target: 67P/Churyumov-Gerasimenko

Masateru Ishiguro[1]

[1] Department of Physics and Astronomy, Seoul National University, Seoul 151-742, Korea

ishiguro@astro.snu.ac.kr

Running head: Dust Trail of Comet 67P/Churyumov-Gerasimenko

Editorial correspondence to:

Masateru Ishiguro, Dr.

Astronomy Program, Department of Physics & Astronomy,

Seoul National University,

San 56-1, Sillim-dong, Gwanak-gu, Seoul 151-742, Korea

E-mail: ishiguro@astro.snu.ac.kr,

Tel: +82-2-880-6754, Fax: +82-2-887-1435

**Abstract**

A thin, bright dust cloud, which is associated with the *Rosetta* mission target object (67P/Churyumov–Gerasimenko), was observed after the 2002 perihelion passage. The neckline structure or dust trail nature of this cloud is controversial. In this paper, we definitively identify the dust trail and the neckline structure using a wide-field CCD camera attached to the Kiso 1.05-m Schmidt telescope. The dust trail of 67P/Churyumov–Gerasimenko was evident as scattered sunlight in all images taken between September 9, 2002 and February 1, 2003, whereas the neckline structure became obvious only after late 2002.

We compared our images with a semi-analytical dynamic model of dust grains emitted from the nucleus. A fading of the surface brightness of the dust trail near the nucleus enabled us to determine the typical maximum size of the grains. Assuming spherical compact particles with a mass density of $10^3$ kg m$^{-3}$ and an albedo of 0.04, we deduced that the maximum diameter of the dust particles was approximately 1 cm. We found that the mass-loss rate of the comet at the perihelion was $180 \pm 50$ kg s$^{-1}$ on or before the 1996 apparition, while the mass-loss rate averaged over the orbit reached $20 \pm 6$ kg s$^{-1}$. The result is consistent with the studies of the dust cloud emitted in the 2002/2003 return. Therefore, we can infer that the activity of 67P/Churyumov–Gerasimenko has showed no major change over the past dozen years or so, and the largest grains are cyclically injected into the dust tube lying along the cometary orbit.

# 1. Introduction

The Jupiter family comet 67P/Churyumov–Gerasimenko (67P/C-G) is the target of the European Space Agency (ESA) *Rosetta* space mission. Due to the delay of the *Rosetta* mission, ESA changed the mission target from 46P/Wirtanen to 67P/C-G in early 2003 while 67P/C-G was moving away from the sun after its perihelion passage on August 18, 2002. *Rosetta* will rendezvous with 67P/C-G in May 2014 at a heliocentric distance of 4.5 AU, about 15 months before its perihelion passage. Because of its dust trail, 67P/C-G is known as a comet that emits large dust particles (Sykes and Walker, 1992). The large dust grains greater than 1 mm are not sensitive to solar radiation pressure and may remain in the vicinity of the nucleus, thus posing a potentially serious impact on the spacecraft. Therefore, it is important to study the dust environment, especially the large grains ejected from 67P/C-G.

Large dust grains generally remain close to the orbit of their parent bodies for many revolutions around the Sun, and form the dust tubes or trails similar to airplane contrails. Neckline structures are temporal brightness enhancements in the dust trail caused by the particles ejected at the point (first node) $\pi$ away in true anomaly from the observed point (second node) (Kimura and Liu, 1977; Fulle and Sedmak, 1988). The initial spherical shell that is ejected isotropically from the nucleus before the perihelion becomes ellipsoidal in shape by collapsing on the orbital plane of the parent comet at the second node. As a result, the shell appears as a narrow, extended or neckline structure when the obsever is close to the comet's orbital plane.

In 2003, 67P/C-G's appearance was quite intriguing. The comet exhibited a thin bright dust cloud, which has been the subject of some controversial interpretation. In their initial report, Reach et al. (2003) inferred that the thin dust cloud or dust trail could be composed of dust released around the previous perihelion on January 1996. Fulle et al. (2004) applied their three-dimensional dust dynamic model as well as the neckline model to conclude that the thin bright cloud was composed of the dust particles ejected more recently, in the period approximately 150 days before and after the perihelion. Moreno et al. (2004) used the same approach and argued that the thin cloud is not a dust trail. Therefore, it is likely that the thin bright cloud observed in early 2003 is the neckline structure. Efforts to detect the dust trail are still ongoing. Among these are the concerted efforts of Agarwal et al. (2006, 2007), who barely managed to distinguish the dust trail from the neckline at the heliocentric distance of 4.7 AU. However, since the angular separation between the neckline and the dust trail was $0.8°$ in their image, this was not sufficient to extract the brightness profile of the dust trail, and the obtained flux of the dust trail was highly contaminated by the neckline (Agarwal et al., 2007).

In this paper, we present images of multi-epoch observations of 67P/C-G during 2002 and 2003 apparition using Kiso 1.05-m Schmidt. Although the aperture of the telescope is less than half of those applied for the previous studies (Fulle et al., 2004; Moreno et al., 2004; Agarwal et al., 2006, 2007; Reach et al., 2003), the wide coverage of the sky as well as a careful data reduction enabled to detect a faint structure. In this paper, we focus on the detection and the interpretation of the dust trail associated with 67P/C-G. We explain the detail of the observations and the data reduction in Section 2. Not only the thin bright structure but also faint extended cloud was detected. We interpret the extended structure in Section 3, We describe the model to estimate the mass-loss rate in Section 4. Although the mass-loss rate was deduced from the dust cloud emitted during 2002/2003 apparition (Fulle et al., 2004; Moreno et al., 2004), our approach differs from theirs in that we determine the mass-loss from the "dust trail".

## 2. Imaging and Data Reduction of 67P/C-G

### 2.1. Observations

We conducted charge-coupled device (CCD) observations of 67P/C-G using the Kiso Observatory 1.05-m Schmidt telescope (35°47'39" N, 137°37'42" E, at an elevation of 1130 m at Nagano, Japan) between September 9, 2002 and February 1, 2003. We used the 2KCCD camera, with a field of view of 50' × 50' and a pixel scale of 1.5"/pixel. Since the R-band (effective wavelength of 0.64 μm) of the Kiso Schmidt telescope is the most sensitive to solar-color objects, we used the R-band filter. Individual exposure times were in the range of 3 to 5 min depending on the apparent movement relative to the stars and the sky conditions. The observations, as well as the observational conditions such as the position and the phase angle of the comets, are summarized in Table 1. Although the seeing at the Kiso Observatory are notably bad, this is not important for extended sources like the cometary dust cloud. Since all the images were taken under variable weather conditions, we did not observe the standard stars listed in the photometric standard references such as the Landolt catalog. Instead, the flux calibration was performed using the field stars listed in the USNO-A2.0 catalog (Monet, 1998). Fig. 1 shows the position of the Earth and 67P/C-G in the ecliptic coordinate system. 67P/C-G passed the perihelion at a perihelion distance of $q$ = 1.29 AU on August 18.3, 2002, and all of our observations were carried out after that time. In early 2003, we had the opportunity to look along the orbital plane of 67P/C-G. The separation of the position angles of the neckline structure and the orbit of the nucleus projected on the celestial plane were largest in early 2003.

### 2.2 Data Reduction

The data reduction was carried out mainly by programs developed by us (the algorithm is written below). We also used ready-made software, *SExtractor* and *IRAF* (see below).

At optical wavelengths, it is necessary to pay close attention in the data analysis to subtracting stars in order to find faint dust clouds like dust trails. When we combined individual images by average or even median, after applying the offsets to align comets and trails, most of the sky was contaminated by stars and galaxies (Fig. 2A).   There are several methods for star removal.   One method is to apply a rejection operation during the image combination after offsetting.   Another method is to interpolate the pixel values using the data surrounding the stars.   When thresholding parameters are determined well, these methods seem to produce plausible results.   There remain, however, some pixels with values higher than those of dust trails, and it is those we are concerned here.   Most of these pixels are due to halos of stars, and are caused by the forward scattering of the Earth's atmosphere and internal reflection of the optics. Therefore, stars and their neighboring regions should be clearly identified and excluded from the original data.

We applied the following techniques in this study. First, we made images to align stars, since this is an effective way of detecting faint stars and galaxies. Using these images, stars were automatically detected by *SExtractor* (Bertin and Arnouts, 1996), and their magnitudes were measured with the '*phot*' command in the *IRAF/APPHOT* package for aperture photometry. As Fig. 2(b) shows, typically, stars brighter than the 22nd magnitude were detected by this method using images with approximately 30-min exposures. In many cases, by examining images with an image viewer, we manually identified extended objects and faint stars that *SExtractor* could not recognize because they were close to bright stars. We masked the identified objects to a radius of three to five times the full-width at half-maximum (FWHM). We also masked pixels identified as bad through the images of bias (hot pixels and lines) and flatfielding (pixels with high or low sensitivity) as shown in Fig. 2(c). We combined the masked images with offsets to align the comet, excluding the masked pixels and shifting the background intensity to zero as shown in Fig. 2(d). Since the comet moves relative to the stars, in most cases, it was possible to exclude the masked data and combine the sequence of images. Finally, we obtained an image without stars shown in Fig. 2(d). Note that this procedure does not artificially determine pixel values, as would be achieved by interpolation. In the interpolation method, narrow trails tend to disappear. The detection limits depended on the total exposure time and sky conditions, and were roughly 27.5 mag/arcsec$^2$ (or $9.6 \times 10^{-9}$ W m$^{-2}$ sr$^{-1}$ μm$^{-1}$) for September 9, 2002 through December 2, 2002) and 28.5 mag/arcsec$^2$ (or $3.8 \times 10^{-9}$ W m$^{-2}$ sr$^{-1}$ μm$^{-1}$) for February 1, 2003. Note that the lower limit of the dust cloud brightness discussed in the previous studies is 26.6–26.7 mag/arcsec$^2$ (Fulle et al., 2004; Moreno et al., 2004).

We employed the *IMCOORDS* package in *IRAF* to convert pixel coordinates into celestial coordinates. Using this, we determined parameters for the conversion by comparison with USNO-A2.0 star catalog. The estimated astrometric error is 0.5" (1$\sigma$). Therefore, the error of the position angle becomes 0.02º that is enough to identify the trail and the neck-line structure at the time of our observations.

## 3. Interpretation of the obtained Images: Dust Trail and/or Neck-Line Structure

Fig. 3(a) shows the contrail-like cloud detected in the image taken on September 9, 2002. It extended backward (westward) beyond the camera field of view (i.e. longer than 40'). As noted in the previous studies (Fulle et al., 2004; Moreno et al., 2004), the thin bright cloud was observed in the images obtained on December 2, 2002 and on February 1, 2003. This is shown in Figs. 3(b) and 3(c). We initially believed that the thin structure was a dust trail like that originally detected by Infrared Astronomical Satellite (IRAS) observations (Sykes and Walker, 1992). When we examined the images in more detail, however, we noted other structures in Figs. 3(b) and 3(c). These faint structures are likely associated with 67P/C-G, and we refer to them to as "TR" in Fig. 3. In Fig. 3(b), the thin dust cloud extends beyond the field of view and we observed that the cloud splits in two at a distance of about 10' from the nucleus. In the February 2003 image, we also found an extended structure whose position angle is different from both remnants of stars and the neckline. In Fig. 3(c), the surface brightness of the neckline decreases as the comet–neckline distance increases, while the faint cloud (TR) emerges at a distance of about 5'; its brightness is comparable to that of the neckline at a comet–cloud distance of 15'.

Neckline structures are most prominent when the observer is close to the orbital plane of the comet. Moreover, the brightness increases in the period a hundred days or so after the perihelion passage because of abundant dust grains emitted around the perihelion that forms neckline structures. Therefore, the neckline must be prominent in our data. Note that the necklines are faint before or close to the perihelion because of the weak activity of the comet at the point $\pi$ away in true anomaly (i.e., around the aphelion). The true anomaly $f$ at the time of observation in September 2002 was 19°, which means that the dust grains emitted at 5.22 AU ($f = 199°$) could form the neckline (see Table 1). Therefore, it is reasonable to assume that the surface brightness of the neckline would be faint in the image of September 2002. In contrast, the images of December 2002 and February 2003 were observed 106.6 days and 167.5 days after the perihelion passage, and the column density of particles in the necklines would stand out in the cometary dust clouds.

Fig. 4 shows the calculated position angle of the neckline and the orbital plane of the parent body projected on the sky plane. We compared the observed positions with the calculated positions in Fig. 4. It is clear that the thin bright cloud is the neckline, and the faint extended traces are the dust trail composed of the dust grains ejected before the last perihelion passage. For a reference, we plotted the position angle of the thin bright cloud described by Fulle et al. (2004), who contended they had observed the neckline.

Fig. 5 shows the peak brightness of the dust trail. The peak brightness of the December image is not included due to the contamination from the neckline. We made three key observations. First, the surface brightness of the dust trail in the September image appears uniform in the camera field of view. Second, in the February image, the brightness was faint near the nucleus. Third, the trail may not extend forward in the direction of the comet movement. A similar fading was observed in the dust trail image of comet 22P/Kopff (Ishiguro et al., 2002, 2007). This does not mean that trail particles do not exist. It may be due to uncertainty caused by the overlap with the neckline and the bright coma. We plotted the brightness of the dust cloud extended forward, but were not sure that it was the real trail.

## 4. Dust Trail Model

To better understand the observed morphology of the dust cloud of 67P/C-G, especially the dust trail, we create the simulation image using a Monte Carlo dynamical model. We basically follow the model in Ishiguro et al. (2007).

The orbits of dust grains are mainly determined by the ejection speed and the size (parameterized by $\beta$, the ratio of the solar radiation pressure $F_r$ to the gravitational attraction $F_g$). Assuming that the particles are spherical with radius $a$ (cm) and mass density $\rho$ (kg m$^{-3}$), $\beta$ is defined as

$$\beta \equiv \frac{F_r}{F_g} = \frac{K Q_{pr}}{\rho a}, \qquad (1)$$

where $K = 5.7 \times 10^{-5}$ g cm$^{-2}$ and $Q_{pr}$ is the radiation pressure coefficient averaged over the solar spectrum (Burns et al. 1979). Supposing that particles are compact shape and large compared to the optical wavelength, we can fix $Q_{pr} = 1$. We will discuss the case of dust aggregate later.

We assumed that the dust particles were ejected symmetrically with respect to the Sun-comet axis in a cone—shape jet with half-opening angle $w$. Given that the comet's activity was symmetric with respect to the perihelion and the aphelion, we adopted an empirical function for the ejection terminal velocity of dust particles:

$$v_{ej}(r_h, \beta) = V_0 \left(\frac{a}{a_0}\right)^{-u1} \left(\frac{r_h}{AU}\right)^{-u2}, \qquad (2)$$

where $V_0$ is the reference ejection velocity (m s$^{-1}$) of the particles radius $a_0 = 0.5$ cm at the heliocentric distance $r_h = 1$ AU. $u1$ and $u2$ are the power indices of the $\beta$– and the heliocentric distance $r_h$– dependence of the ejection velocity. We supposed $u1 = 1/2$ and $u2 = 1/2$. A power law size distribution with index $q$ was used. We presumed that the dependence between the dust production rate and the heliocentric distance could be expressed by a simple exponential function with power index $k$. The production rate of the dust particles emitted within a size range of *a-da/2* and *a+da/2* in the time range from *t-dt/2* to *t+dt/2* can be expressed by

$$N(a;t)\,dadt = \begin{cases} N_0 \left(\dfrac{r_h(t)}{\text{AU}}\right)^k \left(\dfrac{a}{a_0}\right)^q dadt & a_{min} \le a \le a_{max} \\ 0 & a < a_{min}, a > a_{max} \end{cases}, \quad (3)$$

where $a_{max}$ and $a_{min}$ are the maximum and minimum particle sizes, respectively. $N_0$ is the reference size of $a_0$=0.5 cm-sized dust particles at $r_h$=1 AU. The modeled comets' images were generated by the Monte Carlo approach for the five unknown parameters (i.e. *k*, *q*, $a_{max}$, $V_0$ and *w*). The position of dust particles at a given time is semi-analytically calculated by rigorously solving the Kepler's equations. The calculated position of the dust particles were filtered in order to meet the observed geometries, and the intensities $I_{MDL}(x,y)$ at CCD coordinate (x, y) were obtained,

$$I_{\text{MDL}}(x,y) = \int_{t_0}^{t_{obs}} \int_{a_{min}}^{a_{max}} F_\odot \left(\dfrac{r_h}{\text{AU}}\right)^{-2} a^2 A_p N_{cal}(a;t)\,dadt \;, \quad (4)$$

where $F_\odot$ =1.60 x $10^3$ (W m$^{-2}$ μm$^{-1}$) is the incident solar flux at the solar distance $r_h$=1 AU in R-band. $t_0$ and $t_{obs}$ are the start time of model simulation and the observation time, respectively. $A_p$ is the modified geometric albedo (Hanner et al., 1981). Here we ignore the phase angle dependency of $A_p$ because the phase angle coverage of our measurements (18º-36º) is too narrow to detect the phase angle dependency (Hanner et al., 1989). If we adopt $A_p$=0.04 and the mass density $\rho = 10^3$ (kg m$^{-3}$), the mass-loss rate can be obtained by integrating the differential size distribution with respect to size *a*,

$$\dot{M}(t) = \begin{cases} N_0 \left(\dfrac{r_h(t)}{2\text{AU}}\right)^k \dfrac{4\pi\rho}{3(4+q)} \left(\left(\dfrac{a_{max}}{\text{cm}}\right)^{4+q} - \left(\dfrac{a_{min}}{\text{cm}}\right)^{4+q}\right) & q \ne -4 \\ N_0 \left(\dfrac{r_h(t)}{2\text{AU}}\right)^k \dfrac{4\pi\rho}{3} \left(\ln\left(\dfrac{a_{max}}{\text{cm}}\right)^{4+q} - \ln\left(\dfrac{a_{min}}{\text{cm}}\right)^{4+q}\right) & q = -4 \end{cases}, \quad (5)$$

The average mass-loss rate of a comet over its orbit can be obtained,

$$\dot{M}_{mean} = \frac{\int_{t_{obs}-T}^{t_{obs}} \dot{M}(t)dt}{T} \quad , \tag{6}$$

where $T$ denotes the orbital period.

## 5. Model Results

Dust ejection started from the aphelion in 1986 and continued for about 2.5 revolutions. We assumed a minimum size of $a_{min}$ = 6 μm because such small particles are rapidly blown out by the solar radiation pressure, and they are hard to evaluate for $Q_{pr}$ in Eq. (1). Assuming that $k$, $q$, $a_{max}$, and $V_0$ are in the ranges –2 to –4, –2.5 to –4.5, 500 μm to 5 cm, and 1 m s$^{-1}$ to 15 m s$^{-1}$, respectively, we searched the best-fit parameters. We tested two cases of cone shapes for $w$ = 45° and 90°. Figs. 6–9 show some sample results for $w$ = 45°. Each panel shows the number of particles per pixel weighted by the cross-sectional area of grains. The field of view of these simulation images is 30' × 30', the same as that in Figs. 3(b) and 3(c). These simulation images were convolved by a Gaussian point spread function for comparison with the observed images.

First, we use the example images in Fig. 6 to examine the dependence on $a_{max}$ assuming an initial condition of $k$ = –3, $q$ = –3.5, and $V_0$ = 4.2 m s$^{-1}$. For $a_{max}$ = 5 cm, the prominent dust trail extends not only backward (westward) but also forward (eastward). As $a_{max}$ decreases, the dust trail becomes disconnected from the nucleus and edges toward the trailing direction (i.e., the right-hand side) due to the lack of large particles, which are less-sensitive to solar radiation pressure. Our observation is consistent with the case of $a_{max}$ = 5 mm. Fig. 7 shows the dependence on size distribution $q$ starting from an initial condition of $a_{max}$ = 5 mm, $k$ = –3, and $V_0$ = 4.2 m s$^{-1}$. As $q$ increases, the dust trail becomes clearer due to the injection of many large particles, whereas smaller values of $q$ cause the neckline to appear fainter. Our observations were consistent with a $q$ value of approximately –3.5. Fig. 8 shows the dependence on the dust production rate $k$ in our simulation image. Here we assumed $a_{max}$ = 5 mm, $V_0$ = 4.2 m s$^{-1}$, and $q$ = –3.5. We found that $k$ does not affect the appearance of the comet appreciably, but a value of $k$ = –2 is unlikely because the neckline appears in the September simulation image due to the high activity around the aphelion. Fig. 9 shows the dependence on the initial velocity $V_0$, assuming $a_{max}$ = 5 mm, $k$ = –3, and $q$ = –3.5. As $V_0$ increases, the dust cloud spreads wider and stretches forward. The images we captured are consistent with the simulation images for a value of $V_0$ in the range of 4 to 9 m s$^{-1}$. The estimated ejection velocity at approximately 2 AU agrees closely with the prediction of the neckline model (Fulle et al., 2004; Moreno et al., 2004). If we perform this analysis again for $w = 90º$, we note that the larger value of $w$ increases the width of the dust cloud, especially the dust cloud composed of particles released in the current return, but otherwise, the results are not very different.

The parameter set that can plausibly reproduce the images we captured is as follows: $a_{max}$ ~5 mm, $k \leq -3$, $q$ ~$-3.5$, and $V_0 = 4$–$9$ m s$^{-1}$. We compared the model intensity obtained from Eq. (4) with our observations and focused on the peak brightness of the dust trail in Fig. 10 to determine the mass-loss rate. From Eq. (5), the mass-loss rate at the heliocentric distance $r_h$ is $(400 \pm 120) \times (r_h/\text{AU})^{-3}$ kg s$^{-1}$. We consider only the error originating from the calibration of observed flux by USNO-2.0A.

## 6. Discussion

An image inversion is useful for estimating the size and the ejection speed at a given time. The Finson–Probstein model (Finson and Probstein, 1968) has been widely applied in the interpretation of the dust cloud. It is the approximation that converts the two-dimensional synchrone–syndynes model into the three-dimensional model assuming that the width of synchronic tube is given by the dust ejection speed at the synchrone time. Fulle (1989) developed a sophisticated inversion model that computes the rigorous Keplerian orbits of dust grains. Using Fulle's (1989) method, the dust cloud image of 67P/C-G observed in late March 2003 was studied independently by Fulle et al. (2004) and Moreno et al. (2004). They obtained three major results. First, the mass-loss rate dropped from ~200 kg s$^{-1}$ before the perihelion to ~10 kg s$^{-1}$ after the perihelion. Second, the power index $q$ of the differential dust size distribution at the position of the nucleus dropped from –3.4 to –4.5 through the perihelion passage. Third, the reference ejection speed for a radius of $a_0 = 5$ mm was in the range of 2 to 7 m s$^{-1}$. Our approach differs from the others in that we consider the past comet activity over the last one or two orbital periods of 6 to 17 years. Despite this difference, however, the other studies yielded similar results for the power of size distribution $q$ and ejection speed $v_{ej}$. The consistency suggests that the activity of 67P/C-G is periodic, and has experienced no major change over the past decade or so. We acknowledge that Fulle's (1989) approach most accurately replicated the fresh dust cloud including the neckline structure by focusing on comet activities within 150 days before and after the perihelion in 2002 (i.e., $r_h < 2.1$ AU). However, it seems that the minimum $\beta$ of the particles was determined more accurately by our approach in which we deduced that $\beta_{max} = 10^{-4}$.

Note, however, that we estimated the size based on the assumption of compact particles. Köhler et al. (2007) studied the radiation pressure force on dust aggregates consisting of spherical monomers 0.1 μm in radius, and found the mass of compact particles is underestimated unless the aggregates are more compact than Ballistic Particle–Cluster Aggregates (BPCA). In the case of BPCA particles of amorphous carbon, the maximum size of our estimate reaches ~6 g, which is one order of magnitude larger than our estimate. Therefore, the impact by dust grains up to 10 g should be considered during the mission phase of *Rosetta*.

Using deep imaging data from the European Organization for Astronomical Research in the Southern Hemisphere (ESO) 2.2-m telescope of nearly 10 h integration on April 2004, Agarwal et al. (2007) reported that the dust trail of 67P/C-G was barely distinguishable from the neckline at a heliocentric distance of 4.7 AU. The difference in position angle between the neckline and the dust trail was 0.8º. Therefore, the detected dust trail seemed to be highly contaminated by the neckline because the width of the dust trail and the neckline were larger than the separation. They found the surface brightness of the dust trail was uniform beyond 5' from the nucleus, whereas the brightness of the neckline decreased steeply within 5' of the nucleus. Their simulation predicted that the bright peak around the nucleus was likely caused by millimeter- to centimeter-size particles emitted in the 2002/2003 apparition (i.e., the neckline). Their observational approach is absolutely different from ours in that it concentrated on observations far away from the Sun, where the small particles exhibit little influence. The extrapolation of our model for parameters $a_{max}$ = 5 mm, $q$ = –3.5, $k$ = 3, and $V_0$ = 4.2 m s$^{-1}$ coincides with their observations in three areas. First, the brightness decreases within a distance of 5' from the nucleus. Second, both the neckline and the dust trail extend backward. Third, the brightness of the trail is almost the same as that of the neckline in the region between 15' and 30' from the nucleus.

In our semi-analytical calculation, we presumed that dust ejection started at the aphelion in 1986 and lasted for approximately 2.5 revolutions. We estimated the flux contribution of the dust particles ejected in the 1996 and 1989 apparitions, and found the ratio to be approximately 2:1. To realize the exact contribution of the older particles would require numerical integration of the orbital evolution of dust particles including the planetary perturbations and Poynting–Robertson drag. However, the contribution of the particles released before 1989 would be negligibly smaller than that of particles released after the 1989 perihelion passage. Thus, we can infer that our results are due to the comet activity between 1989 and 1996. The dust trail of 67P/C-G was first found by *IRAS* about 20 years prior to our observations. It is important to compare our results with those of IRAS to clarify the variation in the comet's physical properties as well as the albedo. The extremely faint trail associated with 67P/C-G was detected in 1983 during the HCON1 sky survey. At that time, the trail structure extended ΔMA = 0.1° in front and ΔMA = 1.1° behind (Sykes and Walker, 1992). This led us to consider an explanation of the existence of the forward trail in the IRAS era. As seen in 2P/Encke, the dust trail particles in front are generally explained by big particles much larger than 1 mm ejected at high velocity (Ishiguro et al., 2007). Indeed, our big particle model in Figs. 6(a), (d), and (g), and the high velocity model in Figs. 9(c), (f), and (h), predict that the dust cloud extended forward. We computed the image of 67P/C-G at the time of the IRAS observation using the best-fit parameters of $a_{max}$ = 5 mm, $q$ = –3.5, $k$ = 3, and $V_0$ = 3 m s$^{-1}$, and found that a forward spike appears in our model image. Since the comet was located in the evening sky and the sun–comet vector pointed in the direction of the comet movement, fresh grains would extend forward. It is likely that our model is applicable to the dust activity more than 20 years ago, which suggests that the dust activity of 67P/C-G has been constant for several decades. The dust production rate estimated by *IRAS* was 4 kg s$^{-1}$ (Sykes and Walker, 1992), which is factor of 4 smaller than our results indicated. Three possible reasons may explain this discrepancy. First, the previous study underestimated the total mass because it presumed a grain size of 1 mm, one order of magnitude smaller than ours. This is possible, since Fulle et al. (2004) also noticed their underestimate of size, and corrected the mass of the trail. Second, the albedo of the particles might be larger than the value we used ($A_p$ = 0.04). This is unlikely because an albedo of 0.16 is beyond the nominal value of cometary grains (see e.g. Hanner et al., 1981). Further studies on albedo are expected to be carried out using both optical and infrared space telescopes (Reach et al., 2003; Kelley et al., 2006). Third, the dust production rate in the 1980s may have been smaller than that in the 1990s.. It seems reasonable to assume that the activity of 67P/C-G did not change extensively between the 1980s and 2003 due the first explanation above.

## 6. Summary

In this paper, we analyze the optical images of *Rosetta* mission target 67P/C-G observed by Kiso 1.05cm Schmidt. We focus on the detection and interpretation of the dust trail. Assuming the spherical compact particles, the mass density of $10^3$ kg m$^{-3}$ and the albedo of 0.04, it is found that

1. The maximum size of the particles is about 5 mm in radius,
2. The Mass-loss rate around the perihelion passage is 100 kg s$^{-1}$,
3. Our model results are consistent with the results obtained by study of dust cloud composed of particles released in 2002/2003 apparition (Fulle et al. 2004; Moreno et al. 2004).

Note that both maximum size and mass-loss rate are determined under the assumption of compact particles. In case of dust aggregate, the mass could increases one order of magnitude.

## Acknowledgments

The author expresses his deepest appreciation to the useful discussion and the encouragement of Prof. Seung Soo Hong and Prof. Munetaka Ueno. We also thank Jeong Hyun Pyo for his helpful suggestions on how to improve the model. Fumihiko Usui and Yuki Sarugaku helped observations on September and December 2002, respectively, and all observations were supported by Kiso Observatory, Institute of Astronomy, University of Tokyo.

**Tables**

Table 1. Observational circumstance. $r_h$ and $\Delta$ denote the heliocentric and geocentric distance in AU at the time of observations $t_{OBS}$, $\alpha$ the phase angle in degree, the average seeing size (FWHM, "), $\Delta T$ [days] the time after the perihelion passage, EXP [min] the exposure of each frame, and N the number of images used for the composite images in Figure 3. $r_{NL}$ is the Sun-Comet distances when the particles in the neckline were emitted.

| $t_{OBS}$ [UT] | | $\Delta T$ | EXP | N | Seeing | $r_h$ | $r_{NL}$ | $\Delta$ | $\alpha$ |
|---|---|---|---|---|---|---|---|---|---|
| September 9, 2002 | 18:22-19:18 | +22.5 | 3 | 13 | 4.7 | 1.32 | 5.22 | 1.72 | 36 |
| Secember 2, 2002 | 19:42-20:43 | +106.6 | 3 | 8 | 4.2 | 1.77 | 1.82 | 1.62 | 33 |
| February 1, 2003 | 16:00-19:03 | +167.5 | 5 | 9 | 5.6 | 2.23 | 1.99 | 1.40 | 18 |

**Figure Caption**

Figure 1.

Orbits of 67P/C-G projected on the ecliptic plane (lower-left), the X-Z plane (upper-left) and Y-Z plane (lower-right) of the ecliptic coordinate system. Large ellipse is the orbit of 67P/C-G and small ellipse in the lower left panel is the orbit of the Earth, respectively. ⊙ denotes the position of the Sun and ● means the positions of the comet at the time of the observations. Straight lines are corresponding to the viewing direction from the observer to 67P/C-G.

Figure 2.

Example of data processing for 67P/Churyumov-Gerasimenko obtained on September 9, 2002 (A) Combined image after applying the offsets to align comet without removing stars. (B) Combined image to align stars. Circled stars are automatically detected stars by *SExtractor*. (C) and (D) images masked bad pixels and stars. (E) Resultant image combined masked images after the offsets to align comet.

Figure 3.

R-band (wavelength 0.64 μm) images of 67P/C-G obtained on September 9, 2002 (a), December 2, 2002 (b), and February 1, 2003 (c). These images are the standard orientation in the sky, that is, North is up and East is to the left. The position of the nucleus is marked by "+", and the size of each grid is 5'. The brightness scale limits and the contrast were modulated in order to see the faint extended dust structure. Irregular black and white patches are remnants of stars. To confirm the observed structures we illustrate the calculated positions of the neckline and the dust trail by dashed and solid lines, respectively.

Figure 4.

The calculated position angle of the dust trail (the comet orbit projected on the sky) and the neckline. The observed position angles of the neckline and the dust trail candidates (see the body) are illustrated by the filled circles and crosses. For the reference, the observed position angle of the neckline on March 27.0, 2003 (Fulle et al. 2004) is also plotted in the graph.

Figure 5.
Peak brightness of dust trail observed on September 9, 2002 and February 1, 2003. Since the brightness of the fresh dust cloud is much brighter than that of dust trail, we could not measure the trail brightness at the distance <10'.

Figure 6.
Model images of dust emitted since the 1986 aphelion during ~2.5 revolutions for the observation of September 9, 2002 (a-c), December 2, 2002 (d-f) and February 1, 2003 (g-i). We fix $k=-3$, $q=-3.5$ and $V_0=4.2$ m s$^{-1}$, $w=45°$, and the maximum size of the particle $a_{max}$ is the variable, $a_{max}=5$ cm (a,d,g), 5 mm (b,e,h), and 500 μm (c,f,i). For each panel in this figure and Figs. 7-9, the intensity is proportional to the linear of the flux from the particles. The field of view of Figs. 6-9 are 30' x 30' with celestial coordinate (N upward and E to the left), which are the same as those in Figs. 3b-c.

Figure 7.
Model images of dust emitted since the 1986 aphelion during ~2.5 revolutions for the observation of September 9, 2002 (a-c), December 2, 2002 (d-f) and February 1, 2003 (g-i). We fix $a_{max}=5$ mm, $k=-3$, and $V_0=4.2$ m s$^{-1}$, $w=45°$, and the power of dust size distribution $q$ is the variable, that is, $q=-3.0$ (a,d,g), -3.5 (b,e,h), and -4.0 (c,f,i).

Figure 8.
Model images of dust emitted since the 1986 aphelion during ~2.5 revolutions for the observation of September 9, 2002 (a-c), December 2, 2002 (d-f) and February 1, 2003 (g-i). We fix $a_{max}=5$ mm, $q=-3.5$, $V_0=4.2$ m s$^{-1}$, $w=45°$, and the power index of the dust production rate $k$ is the variable, that is, $k$=-2 (a,d,g), -3 (b,e,h), and -4 (c,f,i).

Figure 9.

Model images of dust emitted since the 1986 aphelion during ~2.5 revolutions for the observation of September 9, 2002 (a-c), December 2, 2002 (d-f) and February 1, 2003 (g-i). We fix $a_{max}$=5 mm, $k$=-3, $q$=-3.5, $w$=45º, and the reference ejection velocity $V_0$ (ejection speed of the particles radius $a_0$=0.5 cm at heliocentric distance $r_h$ = 2 AU) is the variable, that is, $V_0$=4.2 m s$^{-1}$ (a,d,g), 8.4 m s$^{-1}$ (b,e,h), and 12.6 m s$^{-1}$ (c,f,i).

Figure 10.

Peak brightness of dust trail fitted by the dust trail model.

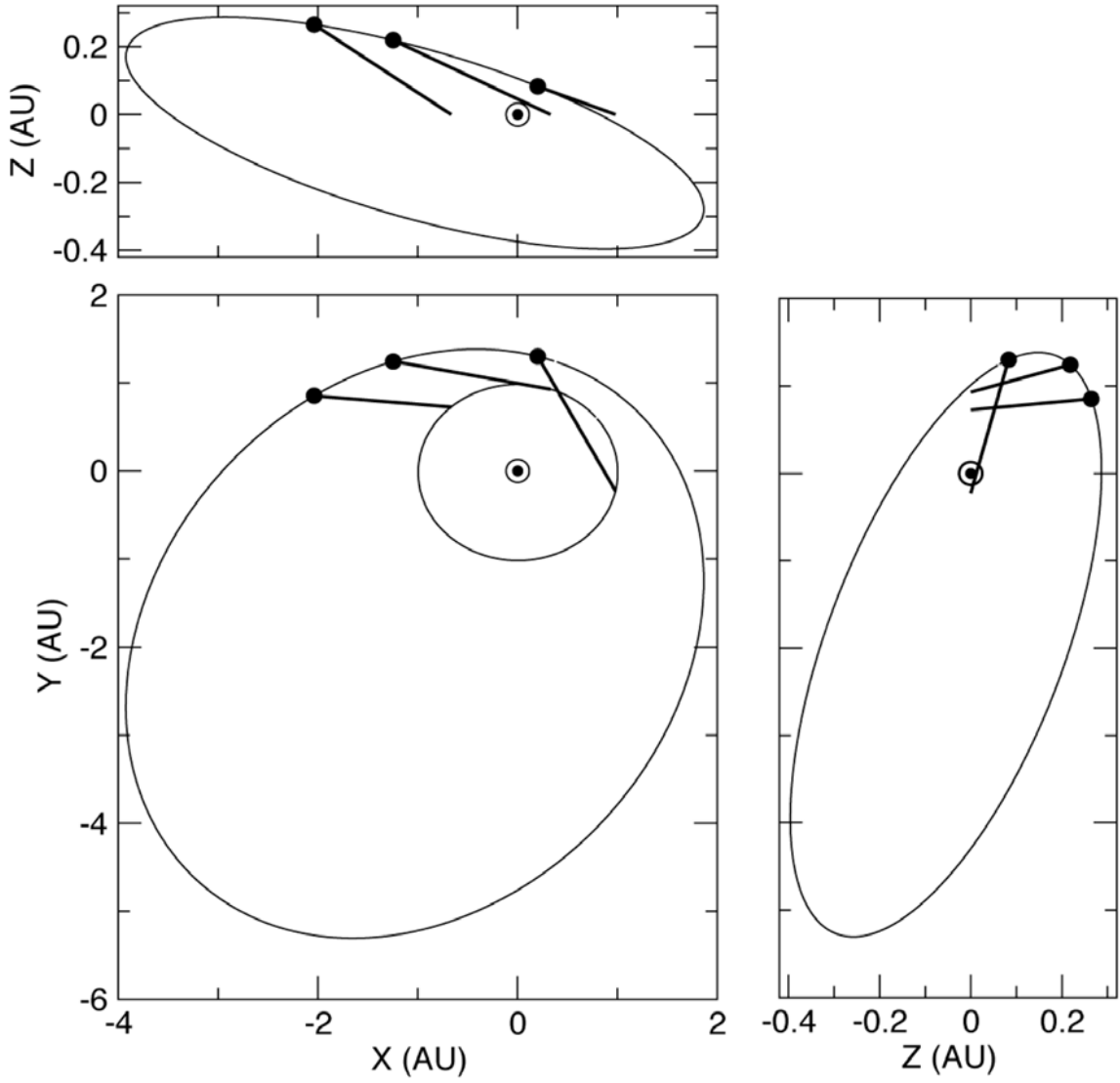

Figure 1. by M. Ishiguro

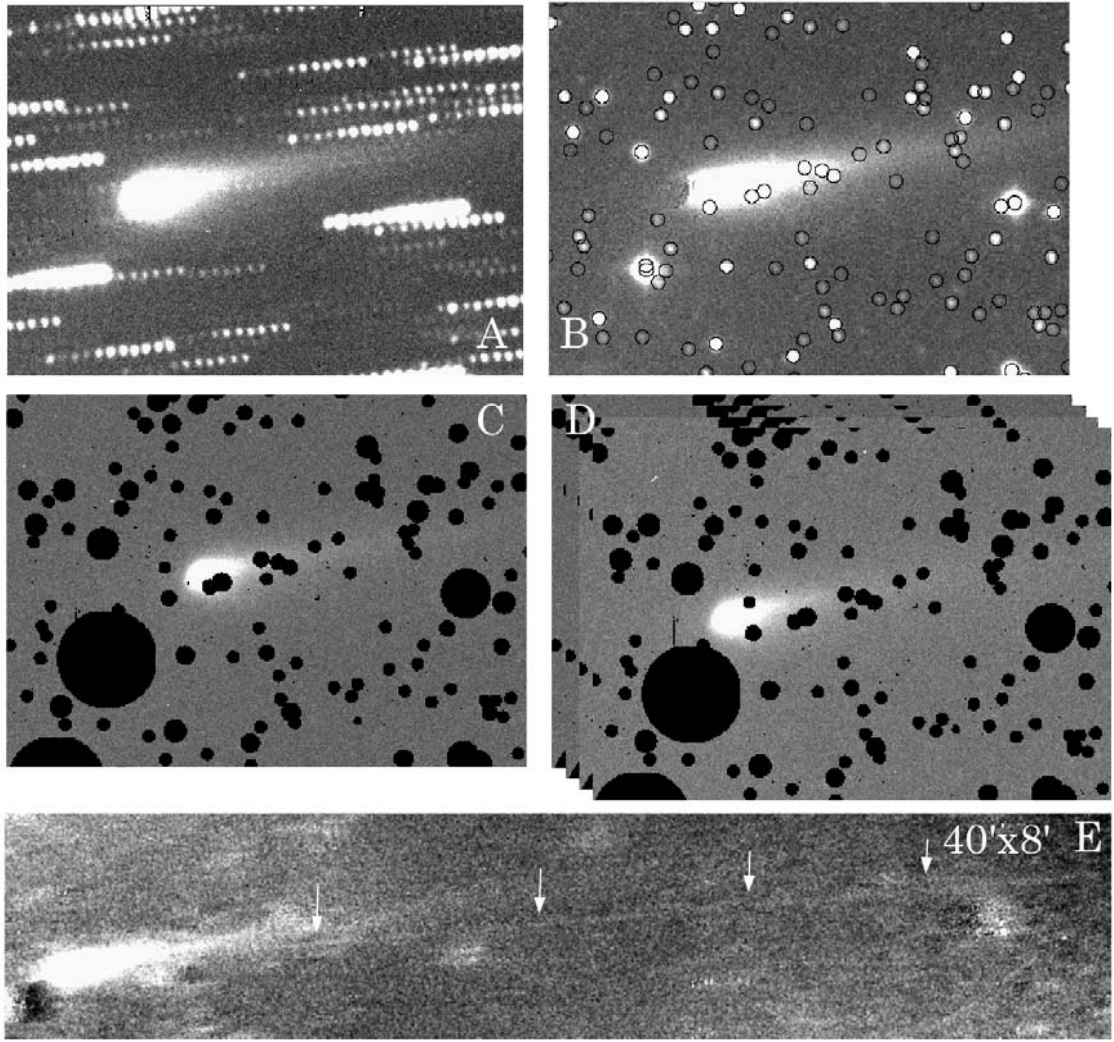

Figure 2. by M. Ishiguro

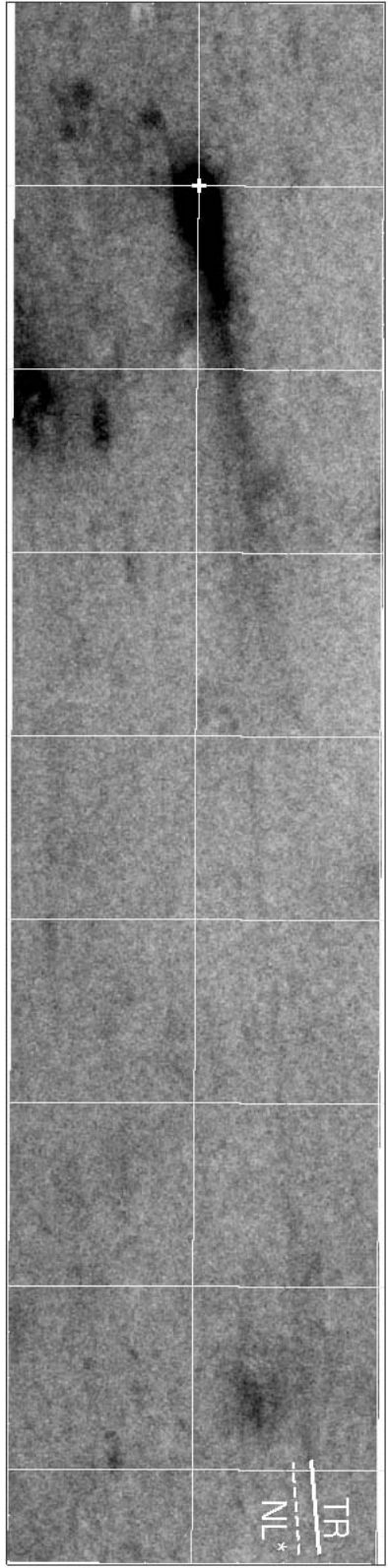

Figure 3a. by M. Ishiguro

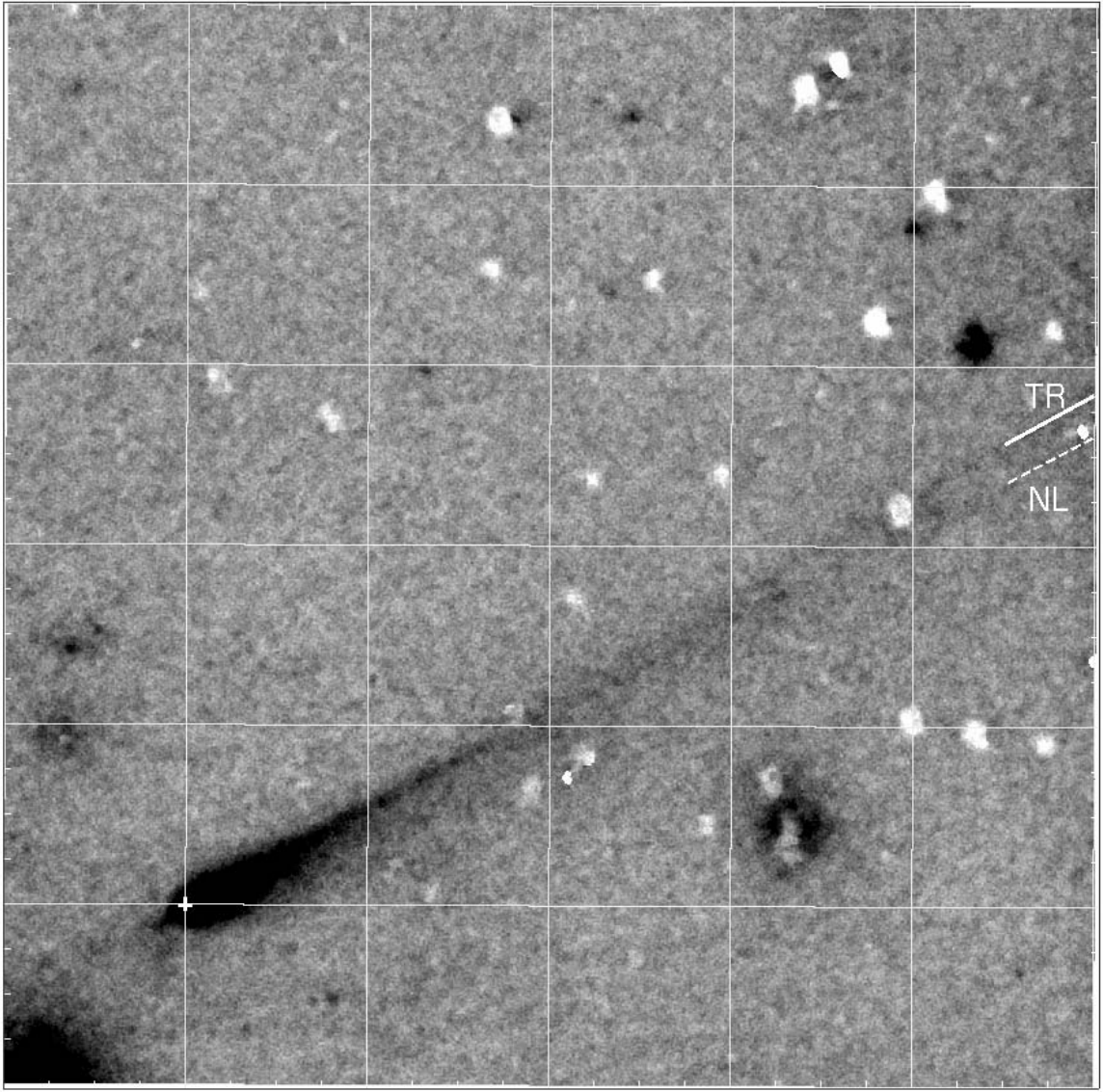

Figure 3b. by M. Ishiguro

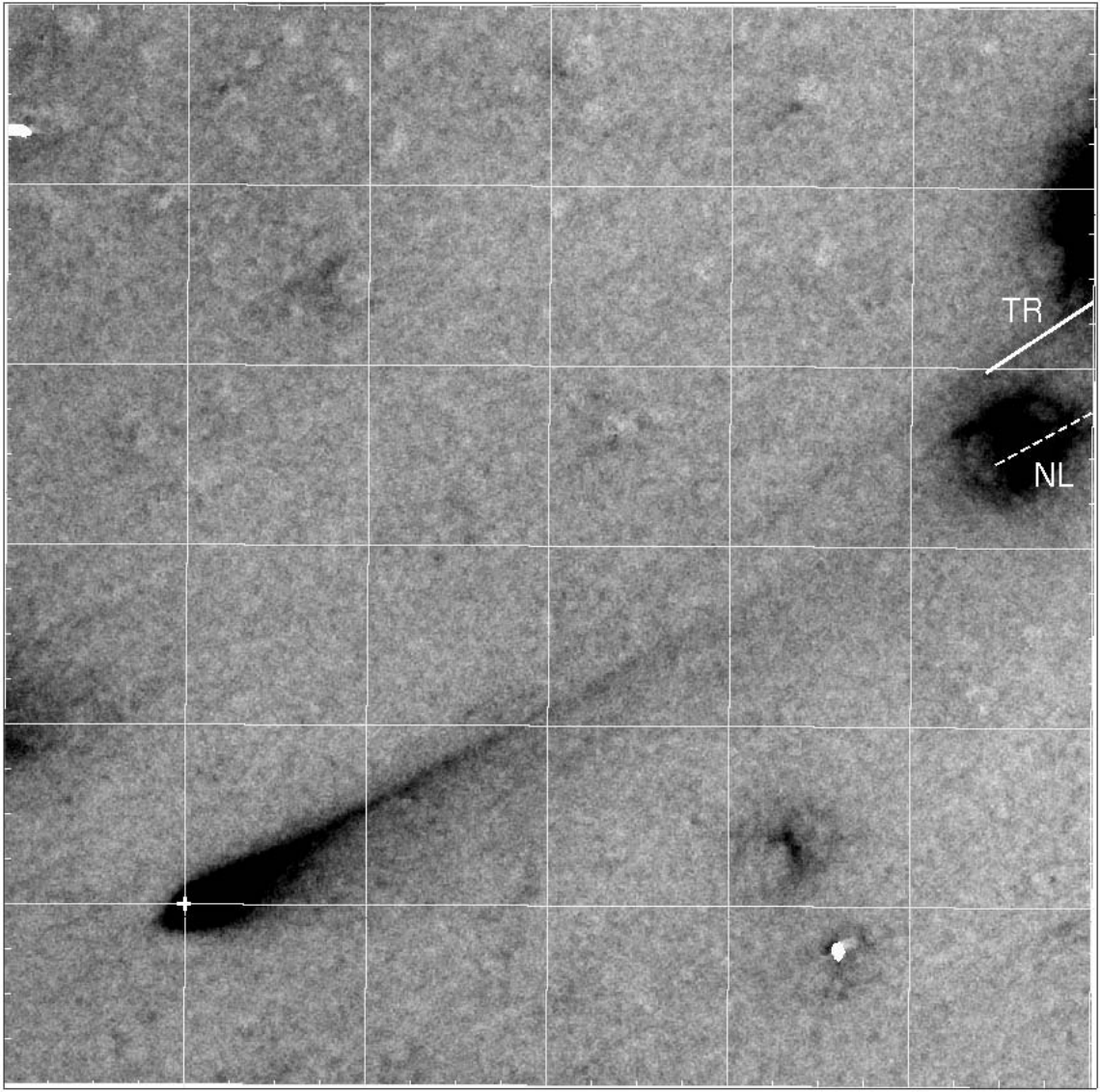

Figure 3c. by M. Ishiguro

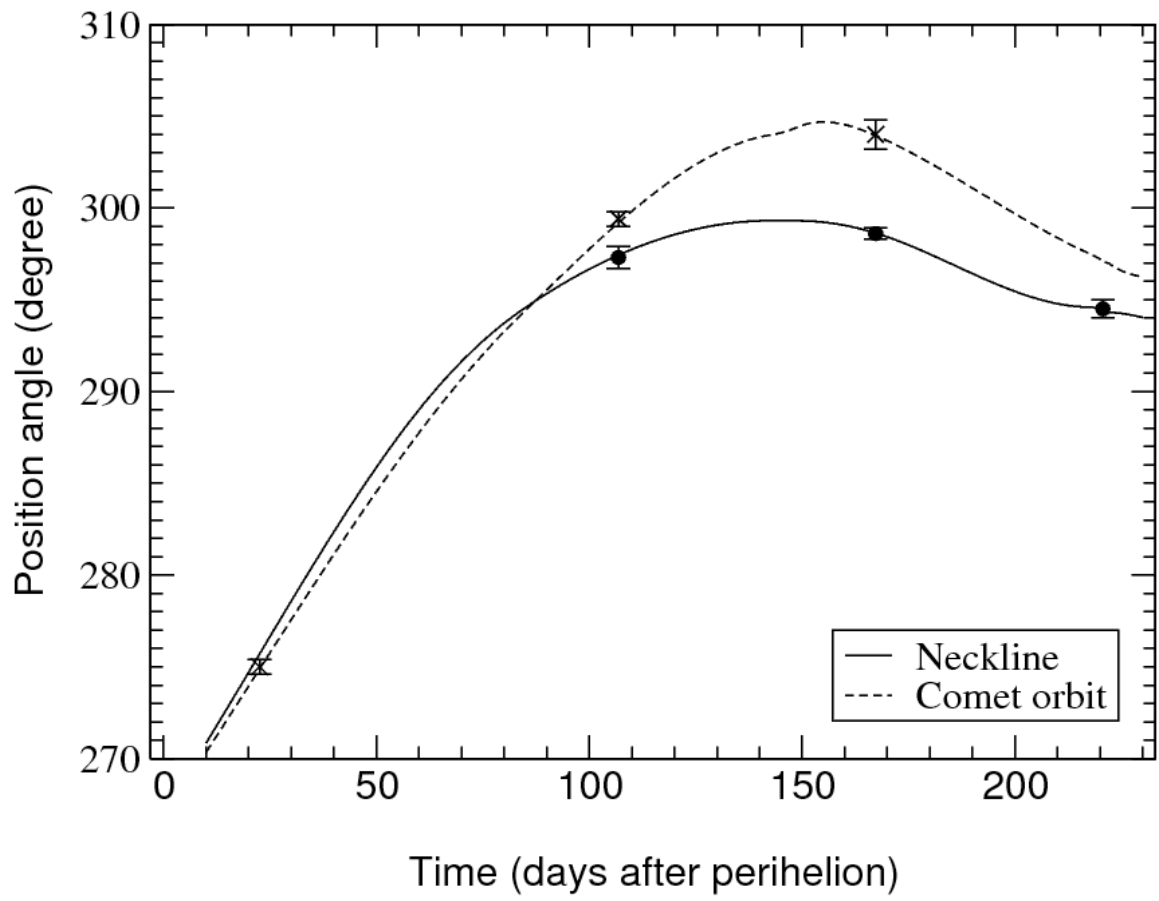

Figure 4. by M. Ishiguro

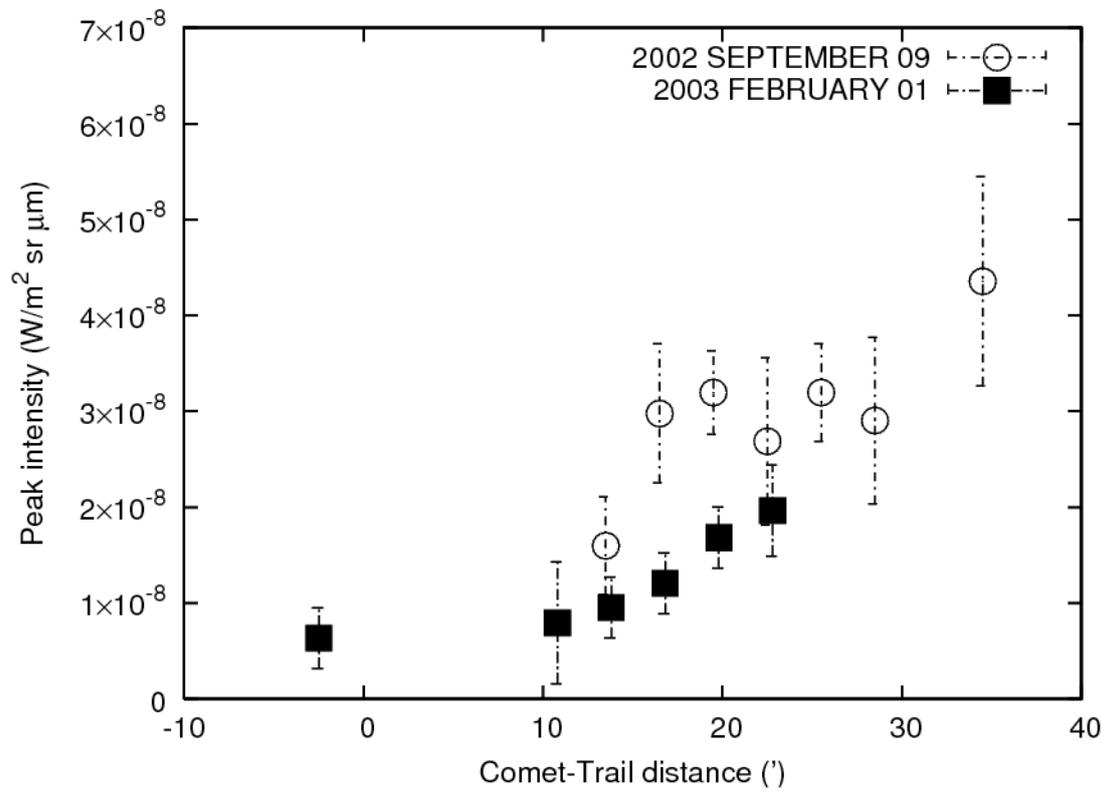

Figure 5. by M. Ishiguro

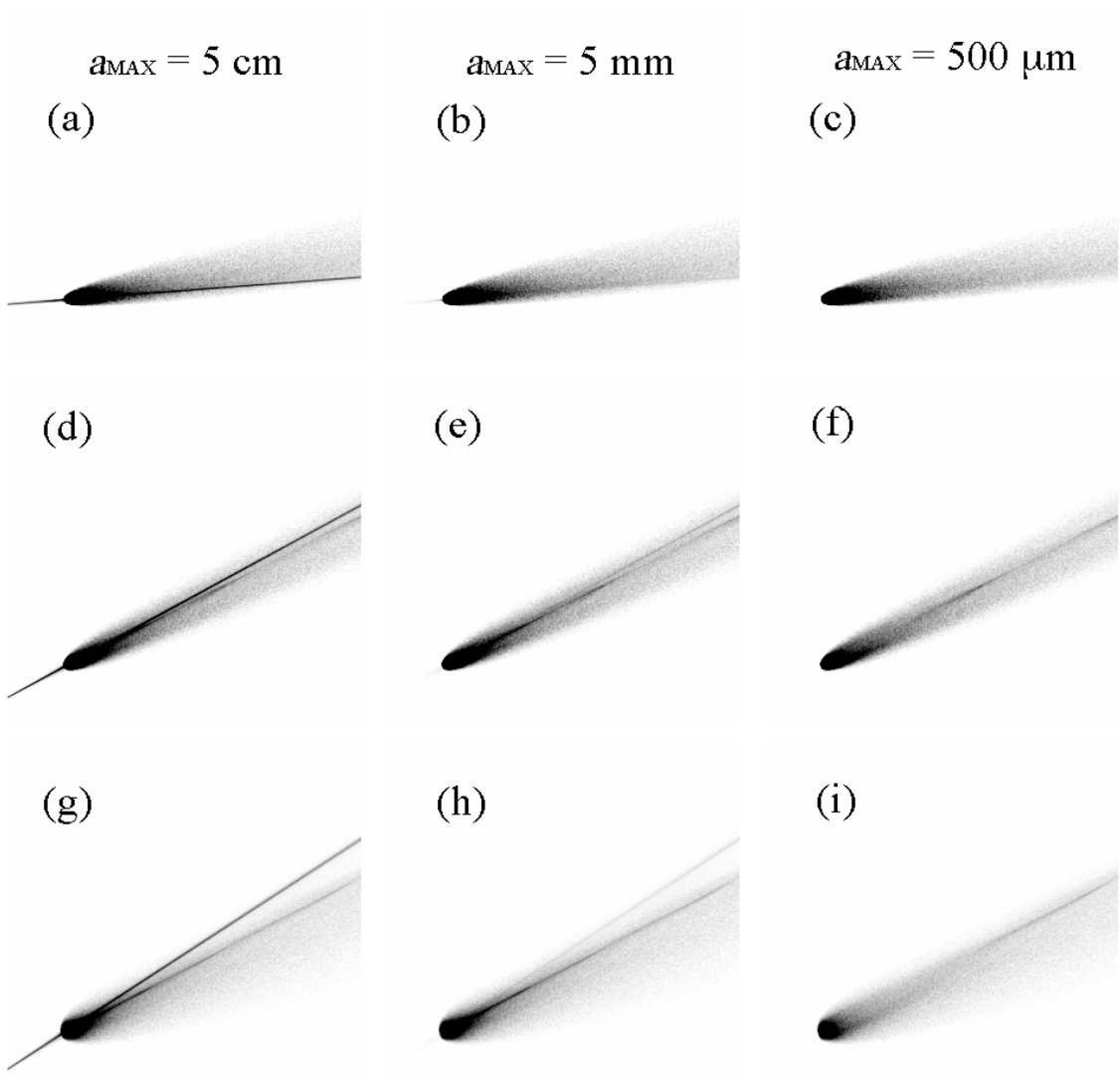

Figure 6. by M. Ishiguro

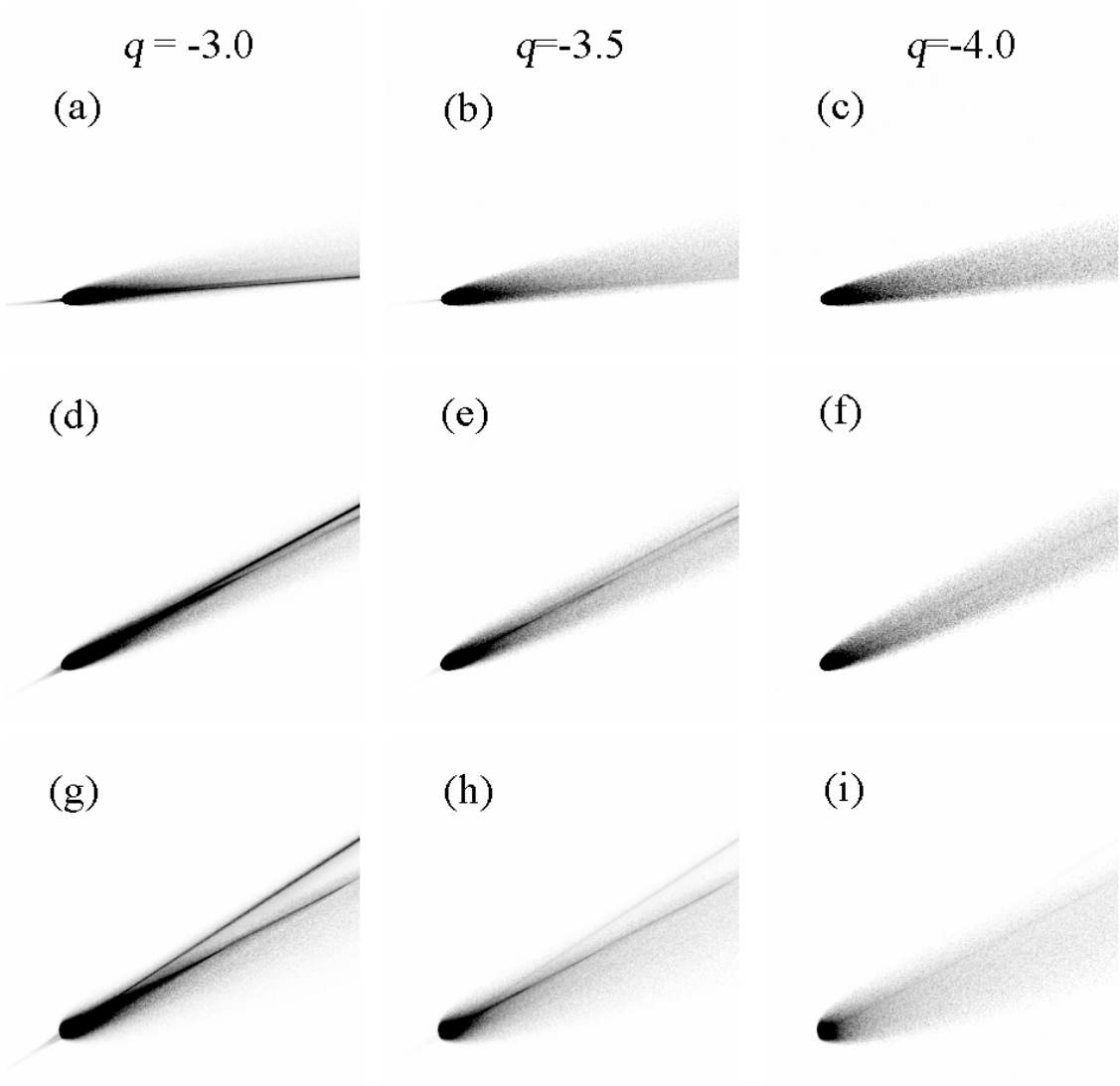

Figure 7. by M. Ishiguro

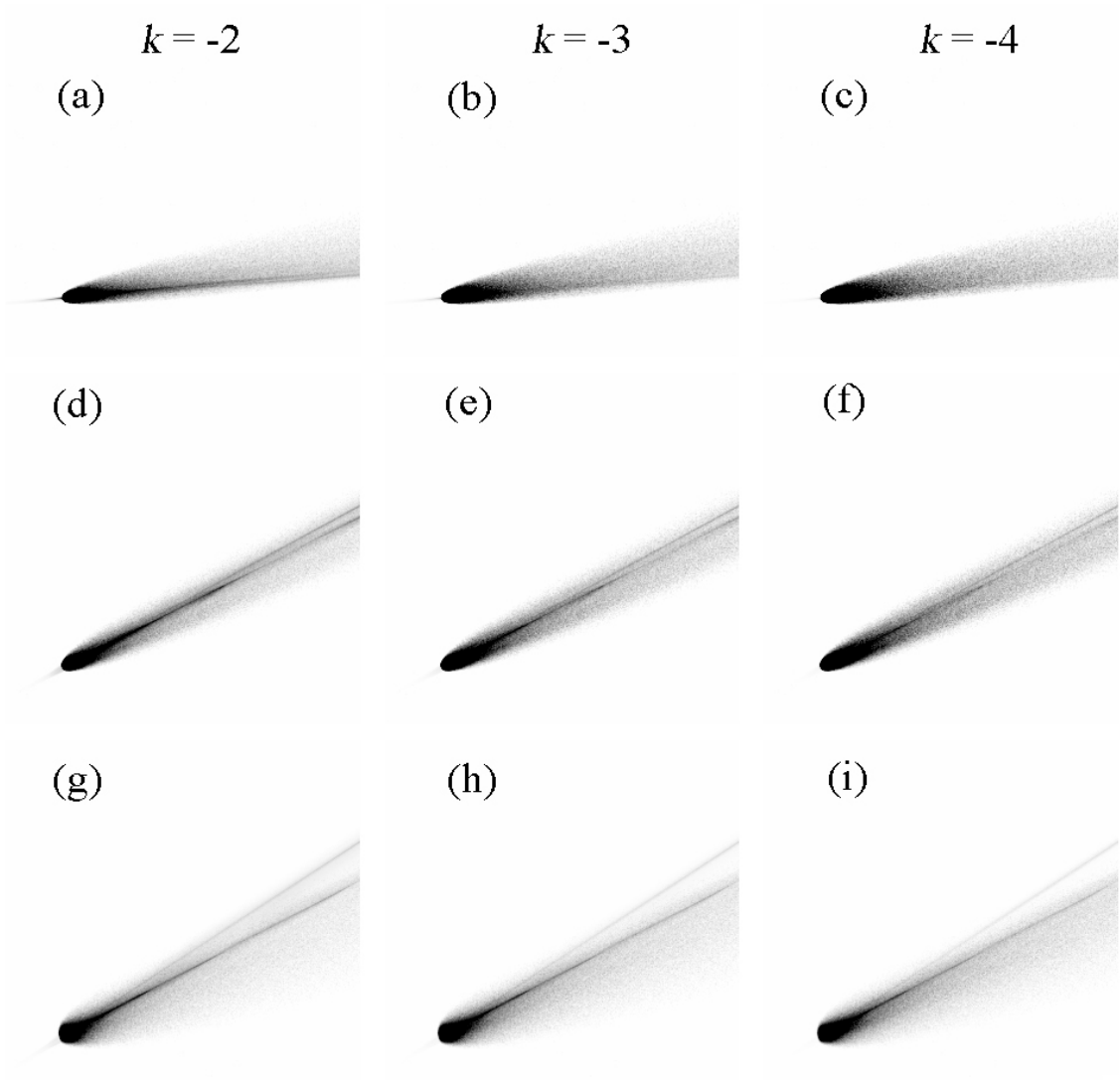

Figure 8. by M. Ishiguro

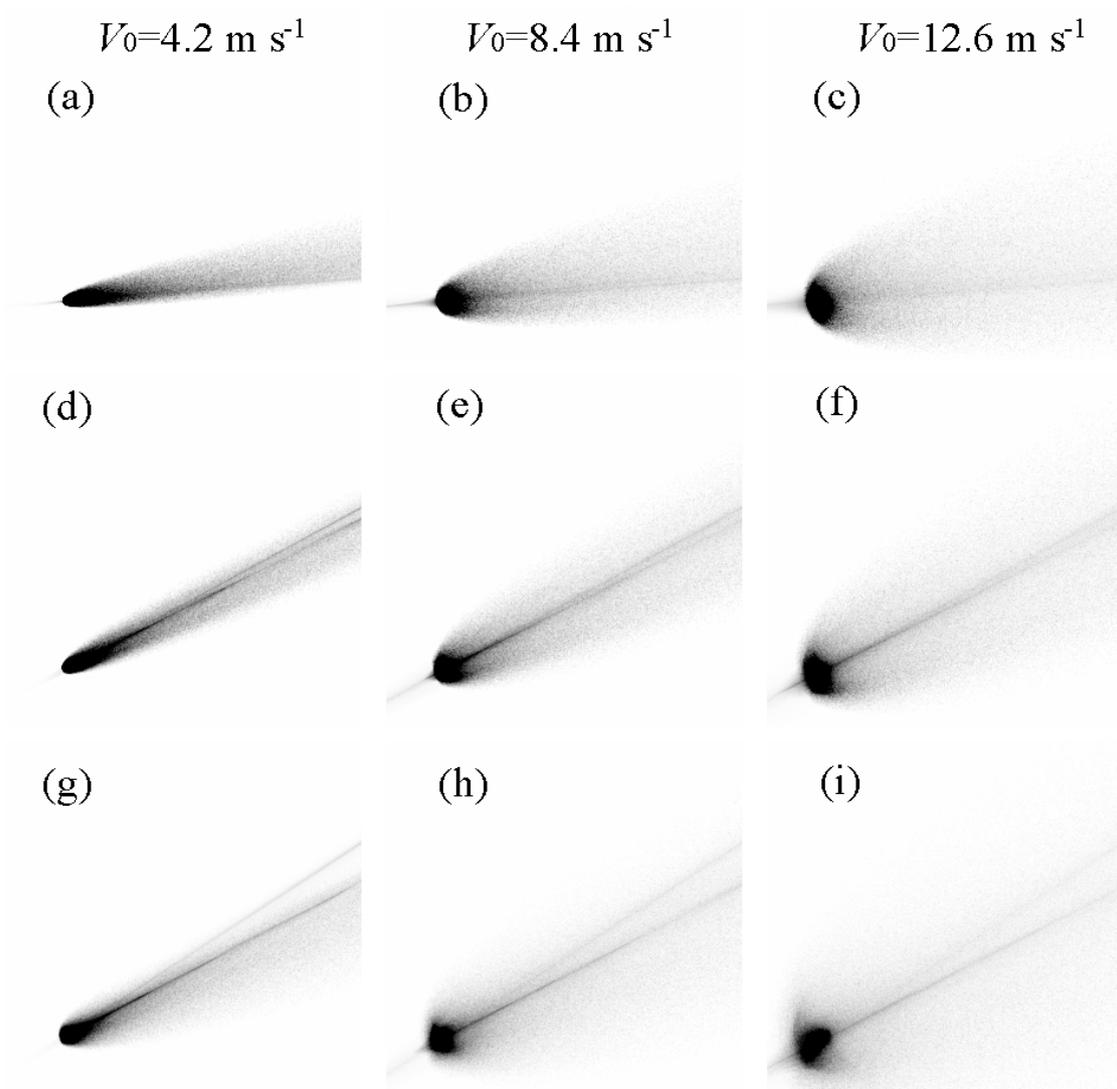

Figure 9. by M. Ishiguro

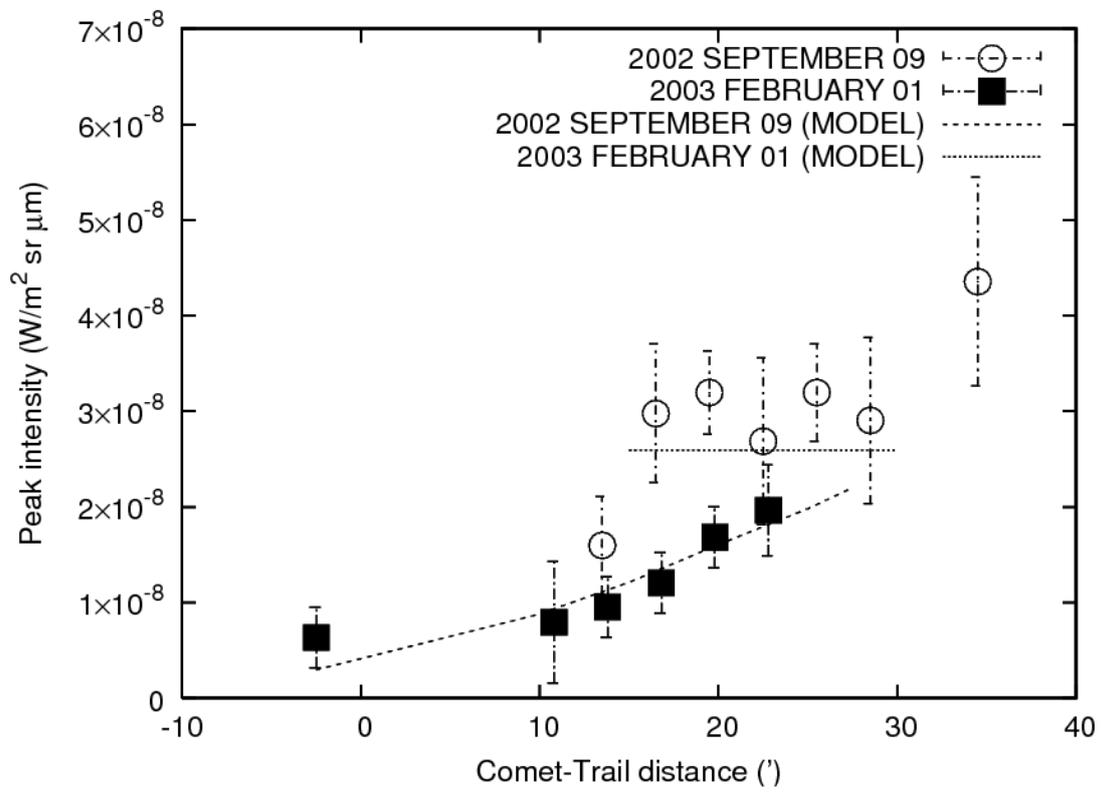

Figure 10. by M. Ishiguro

(END OF DRAFT)